# Layer-tunable Hubbard bands probed via moiré excitons in $MoSe_2/WS_2$ heterostructures

Hongyu Yao[1], Qiao Li[1], Chih-En Hsu[2,3], Takashi Taniguchi[4], Kenji Watanabe[5], Hung-Chung Hsueh[3], Zhenglu Li[2], Andrew Y. Joe[1,*]

[1] Department of Physics and Astronomy, University of California, Riverside, CA 92521, USA.
[2] Mork Family Department of Chemical Engineering and Materials Science, University of Southern California, Los Angeles, CA 90089, USA
[3] Department of Physics, Tamkang University, Tamsui, New Taipei 251301, Taiwan
[4] Research Center for Materials Nanoarchitectonics, National Institute for Materials Science, 1-1 Namiki, Tsukuba 305-0044, Japan.
[5] Research Center for Functional Materials, National Institute for Materials Science, 1-1 Namiki, Tsukuba 305-0044, Japan.
* Corresponding author. Email: andrew.joe@ucr.edu

## Abstract

Moiré superlattices in transition metal dichalcogenide heterostructures provide a highly tunable platform for engineering strongly interacting states at the nanoscale. However, quantitatively determining and in-situ tuning of the underlying Hubbard parameters remains experimentally challenging. Here, we report electric-field-driven reordering of layer-specific Hubbard bands by performing optical spectroscopy on a dual-gated, 60°-aligned $MoSe_2/WS_2$ heterobilayer. Using two spatially distinct moiré excitons as local optical probes and tracking them as a function of carrier filling and vertical electric field, we quantitatively extract the layer-dependent on-site Coulomb repulsions, $U_M \approx 60$ meV in $MoSe_2$ and $U_W \approx 30$ meV in $WS_2$. Furthermore, we stabilize generalized Wigner crystal and stripe phases by electrostatically tuning the system to a type-II band alignment, shifting the ground state into the $WS_2$ layer where reduced on-site repulsion allows inter-site Coulomb interactions to dominate. Our results establish vertical electric fields as a deterministic tuning knob for layer-selective Hubbard physics, enabling device-level control of complex many-body phases.

## Introduction

Moiré superlattices provide a highly tunable platform for realizing strongly correlated interaction-driven quantum phases through the formation of spatially localized electronic states with enhanced Coulomb interactions. These superlattices can be engineered by vertically stacking two-dimensional (2D) van der Waals (vdW) heterostructures with a controlled twist angle or lattice mismatch. In particular, transition metal dichalcogenide (TMD) moiré heterostructures offer an exceptional opportunity to realize correlated electronic states owing to their strong electronic confinement[1–7], electrically tunable band alignment[8–13], and highly controllable interaction landscape[14–18]. Recent experiments have revealed a wide range of correlated phases in these systems, including Mott insulating states[4,19–23], generalized Wigner crystals[19,21,23], and correlated magnetic phases[17,24–26]. The correlated electron behavior in these phases can be optically probed by moiré excitons[1,6,27], electron-hole bound pairs trapped within the periodic moiré potential that are highly sensitive to the local electronic configuration and dielectric environment[3,6,7,19,20,23,27–30]. Although previous studies have shown that excitonic responses can sensitively track correlated electronic phases, how layer-dependent Hubbard interactions govern the formation of these states and how the electric-field-driven reordering of layer-specific Hubbard bands manifests in the excitonic response remains largely unexplored.

Here, we report the observation of correlated insulating states and the electric-field tunability of layer-specific Hubbard bands by probing two distinct moiré excitons in near 60-degree (H-type) $MoSe_2/WS_2$ moiré heterostructures. Importantly, the unique interlayer conduction band alignment in $MoSe_2/WS_2$ moiré superlattice enables selective control over the Hubbard band filling configuration for filling factors 1 to 4, allowing us to quantitatively extract the layer-dependent on-site Coulomb interaction Hubbard $U$, yielding $U_M \approx 60$ meV in $MoSe_2$ and $U_W \approx 30$ meV in $WS_2$. Furthermore, by tuning the heterostructure into a type-II band alignment, we observe generalized Wigner crystal and stripe phases, demonstrating the dominance of inter-site Coulomb interactions when electrons fill the lower Hubbard band of the $WS_2$ layer. Our work establishes an electric-field driven layer-tunable Hubbard platform, providing a direct route toward deterministic control of Hubbard band hierarchy and correlation-driven quantum states in moiré materials.

## The Intrinsic Moiré Exciton Landscape

Figure 1a shows the schematic and electrical circuit of the $MoSe_2/WS_2$ moiré heterobilayer device. We align the $MoSe_2$ and $WS_2$ layers using their optically determined edges and encapsulate the device with hexagonal boron nitride (hBN). Few-layered graphite is used for the top and bottom gates as well as the electrical contact to the TMD layers. Independent control of the top ($V_{tg}$) and bottom ($V_{bg}$) gate voltages allows for full control over the out-of-plane electric field ($E$) and the electrostatically induced carrier density ($n$). The natural lattice mismatch of approximately 4.4% and the relative twist angle between the $MoSe_2$ and $WS_2$ layers jointly creates a moiré superlattice with a periodicity of around 8-10 nm (see Method: Determination of moiré density and filling factor). In this study, we focus on H-type (near 60-degree aligned) heterostructures, specifically Device A (see Extended Data Figure 1 for Device B), which has a twist angle of $\theta \sim 58.4$ degrees. An optical image of Device A is shown in Figure 1b.

We first characterize the neutral excitonic response of the $MoSe_2/WS_2$ heterobilayer using differential reflectance spectroscopy. In Figure 1c, we show a representative differential reflection spectrum ($\Delta R/R_0$). We observe three primary excitonic resonances at 1.62 eV, 1.65 eV and 1.73 eV, originating from the monolayer $MoSe_2$ A-exciton, which we label as $X_{M1}$, $X_{M2}$ and $X_{M3}$, respectively[6,27]. We also observe an excitonic response at 2.06 eV ($X_{W1}$) corresponding to the monolayer $WS_2$ A-exciton energy and a resonance at 1.82 eV corresponding to the B-exciton transition in $MoSe_2$[31]. The inset in Figure 1c depicts the initial type-I band alignment between $MoSe_2$ and $WS_2$, indicating the intralayer excitonic nature of these resonances. The observed excitons do not shift with an electric field (Extended Data Figure 2), confirming their intralayer character.

Theoretical calculations of the moiré excitons in H-type $MoSe_2/WS_2$ heterostructures support these assignments (see Methods: First-principles calculations). Our GW-Bethe-Salpeter equation (GW-BSE) calculations of the excitonic spectrum reproduce two strong resonances $X_{M1}$ and $X_{M2}$ (Fig. 1d). Spatial mapping of the corresponding moiré excitons further reveals their distinct real-space characters (Fig. 1e-f). We find that $X_{M1}$ forms a tightly localized exciton, where both its electron and hole are strongly localized at the moiré $B^{S/Se}$ site. In contrast, $X_{M2}$ corresponds to an intralayer charge-transfer exciton[6], in which the electron is primarily localized at the $B^{S/Se}$ site but the hole is localized at the $B^{W/Mo}$ site. Consequently, their distinct spatial extents allow these two moiré excitons to act as site-specific optical probes of the local charge environment.

**Mott insulating states detected by moiré excitons**

We investigate the doping-dependent differential reflectance in the $MoSe_2$ A-exciton energy window at zero electric field ($E$ = 0 V/nm), as shown in Figure 2a. To only add free carriers to the system without applying a net vertical electric field, we apply $V_{bg} = \alpha V_{tg}$, where α is equal to the ratio between the top and bottom dielectric thicknesses (Methods). Under this condition, the effective net doping is parameterized by the total gate voltage, defined as $V_g = V_{tg} + V_{bg}$. Upon injecting electrons into the system ($V_g \sim 1.2$ V), the $X_{M1}$ resonance exhibits an abrupt 32.5 meV redshift, whereas the $X_{M2}$ resonance only shows negligible energy change. As doping increases, the $X_{M1}$ resonance energy remains almost unchanged, while the $X_{M2}$ resonance redshifts continuously until reaching critical gate voltages of 3.5 V and 5.8 V. At these voltages, $X_{M2}$ exhibits enhanced oscillator strength with a small blueshift. Upon reaching $V_g$ = 8.1V, the neutral $X_{M1}$ peak reemerges with the disappearance of the red-shifted $X_{M1}$ feature.

The abrupt spectral evolution at finite gate voltages with equal spacings implies that unique electronic states emerge at specific fillings of the moiré superlattice. The filling interval $\Delta\upsilon = 1$ corresponds to one electron filling per moiré unit cell, suggesting the system can be described with a moiré Hubbard model (Fig. 2b). As the periodic moiré potential drastically quenches the kinetic energy of the charge carriers, flat conduction minibands form in both the $MoSe_2$ ($C_{M1}$) and $WS_2$ ($C_{W1}$) layers. At zero electric field, the system exhibits a type-I band alignment characterized by an intrinsic conduction band offset ($\Delta^0_{CB}$). Because the on-site Coulomb repulsion ($U$) strongly dominates over the bandwidth, each of these single-particle bands split into a lower Hubbard band ($C^L$) and an upper Hubbard band ($C^U$). Consequently, the relative strengths of the layer-specific on-site Coulomb repulsions ($U_M$ in $MoSe_2$ and $U_W$ in $WS_2$) and the initial conduction band offset ($\Delta^0_{CB}$) determine the energetic ordering and subsequent filling of the Hubbard bands.

We attribute the distinct optical signatures of $X_{M1}$ to the layer-specific sequential filling dictated by this moiré Hubbard model. The discrete energy jump of $X_{M1}$ upon initial electron doping signifies the formation of a strongly bound trion ($X_{M1}^-$). Notably, the observed trion binding energy $\Delta E_{trion}$ = 32.5meV is significantly larger than the free trion binding energy in monolayer $MoSe_2$ (~ 20meV)[31]. Because $X_{M1}$ is localized at the $B^{S/Se}$ site, this enhanced binding energy indicates that the first doped electrons preferentially populate the $MoSe_2$ $B^{S/Se}$ moiré sites filling the lower Hubbard band in $MoSe_2$ ($C_{M1}^{L}$). With further doping, the second integer filling state ($\nu$ = 2) does not affect $X_{M1}^{-}$, implying that the subsequent electrons populate the spatially distinct $WS_2$ AB sites corresponding to $C_{W1}^{L}$. Finally, upon reaching $\nu$ = 3, the disappearance of $X_{M1}^{-}$ and reappearance of the neutral $X_{M1}$ is consistent with the filling of the $MoSe_2$ upper Hubbard band ($C_{M1}^{U}$). Double occupancy of the $MoSe_2$ $B^{S/Se}$ site forms a spin-singlet, which prevents the formation of a bound trion state via Pauli blocking. The overall spatial distribution and layer-specific sequential filling order of the first, second, and third electrons are schematically summarized in Figure 2c.

In contrast, $X_{M2}$ exhibits fundamentally different behavior because of its spatially delocalized nature. Instead of forming a strongly bound trion state with the electrons filling the $MoSe_2$ or $WS_2$ Hubbard bands, it acts as a macroscopic dielectric sensor. Between integer fillings, the continuous redshift of $X_{M2}$ reflects the continuous increase in screening from a compressible Fermi-sea. However, at $\nu$ = 1 and 2, free carriers form incompressible Mott insulating states. This reduction in macroscopic dielectric screening restores the strong electron-hole Coulomb interaction of the exciton, evidenced by a significant recovery in the oscillator strength and a slight blueshift of $X_{M2}$. Together, the moiré excitonic behavior establish the qualitative energetic order of the Hubbard bands at zero electric field: $C_{M1}^{L} < C_{W1}^{L} < C_{M1}^{U}$.

To verify the spatial localization of electron filling, we compute the flat bands at the conduction band edge using first-principles density functional theory (DFT) calculations, with the band gap subsequently corrected via the GW approximation. Figures 2d–e show the spatial modulation of the electronic charge density associated with the lowest-energy conduction band states originating from the K valley of $MoSe_2$ (Fig. 2d) and $WS_2$ (Fig. 2e). These states exhibit pronounced site-selective localization within the moiré potential, directly reflecting the underlying site-resolved moiré landscape discussed above. In particular, the $MoSe_2$-derived conduction states are predominantly localized at the $B^{S/Se}$ site, while the $WS_2$-derived states are mainly concentrated at the AB sites, demonstrating a clear correspondence between band character and real-space moiré registry. In addition, the second conduction band of $MoSe_2$ is also found to be primarily localized at the $B^{S/Se}$ site (Extended Data Figure 3). As a result, when this band begins to be populated, the enhanced occupation of the same spatial site leads to a stronger on-site Coulomb interaction due to the prior filling of the lower-energy band. This increased electron–electron repulsion effectively renormalizes the band energy, resulting in an upward shift of the second conduction band.

**Quantitative Extraction of Hubbard Band Energies**

To quantify the energetic ordering of the Mott insulating states, we systematically tune the Hubbard band configuration by applying a vertical electric field at constant filling factors. We set $V_{tg}$ and $V_{bg}$ with opposite polarity but with a fixed voltage offset to maintain a constant doping condition. To account for the differences in the dielectric environment between the hBN and the TMD, we use a corrected electric field, $E_{hs} = \frac{V_{tg} - V_{bg}}{\frac{\varepsilon_{TMD}}{\varepsilon_{hBN}}(t_{top}+t_{bottom})+d}$ (Methods).

Figures 3a-c show the electric field-dependent reflection spectra at ν = 1, 2, and 3, where we identify several critical electric fields with distinct changes in the spectral response (see Methods: Critical Field Extraction). To understand the underlying mechanism, we first examine the ν = 1 state. Above a critical field of $E_2$ = 0.055 V/nm, we observe a strong enhancement of $X_{M1}$ resonance and a reduction of the $X_{M1}^-$ and $X_{M2}$ oscillator strengths. We attribute these changes to the electric field shifting the $WS_2$ lower Hubbard band ($C_{W1}^L$) below the $MoSe_2$ lower Hubbard band ($C_{M1}^L$). Physically, this drives an interlayer charge-transfer of electrons from the $MoSe_2$ $B^{S/Se}$ site to the $WS_2$ AB site. Similarly, we can identify other critical electric fields corresponding to interlayer charge-transfer transitions. For ν = 2, we observe two critical fields: $E_{1'}$ = -0.047 V/nm, marking where $C_{W1}^L$ crosses the $MoSe_2$ upper Hubbard band ($C_{M1}^U$), and $E_3$ = 0.104 V/nm, marking where $C_{W1}^U$ crosses $C_{M1}^L$. Finally, at ν = 3, one additional transition occurs at $E_1$ = 0.008 V/nm, corresponding to the transition between $C_{M1}^U$ and $C_{W1}^U$. We note that some of these transitions are observable in multiple filling factors due to the previously mentioned Pauli blocking effect ($E_2$ in both ν = 2 and 3).

By converting the critical electric-field intervals between these transitions into an energy shift ($\Delta$), we can quantitatively extract the energy separations between the layer-specific Hubbard bands. The out-of-plane electric field shifts the relative energies of the layers by $\Delta = e\Delta_{E_{hs}} d$ where $d \approx 0.6\ nm$ is the effective distance between the $MoSe_2$ and $WS_2$ layers. First, we determine the initial lower Hubbard band offset ($\Delta_{CB}$) by evaluating the field required to tune $C_{W1}^L$ across $C_{M1}^L$, which yields $\Delta_{\mathrm{CB}} = eE_2 d \approx 33$ meV. Next the field difference, $E_2 - E_1'$= 0.102 V/nm, measures the energy required to shift $C_{W1}^L$ across the entire $MoSe_2$ Hubbard gap. This corresponds to the $MoSe_2$ on-site energy: $U_M = \Delta_2 - \Delta_1' \approx 61.2$ meV. This analysis is further verified by the field difference of 0.098 V/nm between $E_3$ and $E_1$. A similar analysis is performed to extract the $WS_2$ on-site energy by using the field required to tune either $C_{M1}^U$ or $C_{M1}^L$ across the $WS_2$ Hubbard gap, which yields $U_W = \Delta_1 - \Delta_{1'} = \Delta_3 - \Delta_2 \approx 31.2$ meV, respectively. From these spectral measurements, we have quantitatively extracted $U_M \approx 60$ meV and $U_W \approx 30$ meV.

This pronounced disparity in the correlation energy provides experimental insight into the spatial profiles of the confined charge carriers. Since the effective on-site Coulomb repulsion scales inversely with the spatial extent of the localized states, the larger Hubbard $U$ implies a stronger confinement potential. The substantially different layer-dependent Hubbard $U$ indicates stronger spatial confinement at the $MoSe_2$ $B^{S/Se}$ site than at the $WS_2$ AB site. We attribute this difference to spatial variations in interlayer coupling, which generate distinct moiré trapping potentials at different moiré sites for the two layers in our heterobilayer (Extended Data Figure 3).

**Phase diagram and electric-field induced correlated states**

The simultaneous electrostatic control of the vertical electric field $E$ and the total carrier density provides a powerful approach for mapping the global landscape of correlated electronic phases in moiré heterostructures. By systematically tracking different moiré exciton resonances, $X_{M1}$ and $X_{M2}$, we construct a comprehensive phase diagram describing the relative Hubbard band ordering and the evolution of correlated states across the full parameter space.

Figures 4a-b show the minimum differential reflectance intensity of $X_{M1}$ and $X_{M2}$ as functions of the calculated electric field $E$ and the combined gate voltage $V_g = V_{tg} + 0.75V_{bg}$. In Extended Data Figure 4, we highlight the energy ranges considered and also show the minimum differential

reflectance intensity of $X_{W1}$, which exhibits similar transitions at critical electric fields. The combined optical response of $X_{M1}$ and $X_{M2}$ enables direct identification of both the layer-resolved Hubbard band hierarchy and the total filling factor $\nu$. Notably, we observe the emergence of a pronounced $X_{M1}$ response above the critical field $E_2$, signaling a transition of the electronic ground state from the $MoSe_2$ layer into the $WS_2$ layer.

In Figure 4c, we construct a complete phase diagram of the layer-dependent Hubbard band ordering and total filling factor ν by combining the $X_{M1}$ and $X_{M2}$ responses. Distinct regions in the phase diagram correspond to different electrically controlled filling sequences of the Hubbard bands. Across the entire parameter space, the experimentally observed phase boundaries are in excellent agreement with the proposed Hubbard band model, demonstrating that the evolution of the correlated states is governed by the electric-field-controlled reordering of the layer-resolved Hubbard bands.

Beyond the overall band reordering, the phase diagram also reveals the emergence of additional correlated features once the lower Hubbard band of $WS_2$ is tuned below that of $MoSe_2$ (the first light green region in Fig. 4c). To investigate this regime in greater detail, we perform doping-dependent measurements at $E = 0.076\,V/nm$, where the $WS_2$ lower Hubbard band is the energetic ground state. Figure 4d shows the corresponding differential reflectance linecut as a function of carrier density. In this regime, the initially injected electrons ($\nu \leq 1$) preferentially occupy the $WS_2$ AB site. Consistent with our previous observations, this regime corresponds to a stronger $X_{M1}$ oscillator strength. More strikingly, $X_{M1}$ exhibits additional pronounced intensity enhancements near fractional fillings of 1/3, 1/2 and 2/3 of the moiré lattice. To clearly highlight these features, we track the minimum differential reflectance intensity of $X_{M1}$ as a function of carrier density (Fig. 4e), which clearly exhibits local minima at these fractional filling factors. These features are absent in regimes where electrons first occupy the $MoSe_2$ layer and only emerge after the electric field drives the lower Hubbard band of $WS_2$ below that of $MoSe_2$. These observations strongly suggest the formation of correlated fractional filling states within the $WS_2$ lower Hubbard band. We attribute the features at 1/3 and 2/3 filling to the formation of generalized Wigner crystal (GWC) states, while the enhancement at 1/2 filling corresponds to the emergence of a stripe phase[32].

We attribute the emergence of these generalized Wigner crystal and stripe phases to the layer-dependent Hubbard interaction energy. Although the standard on-site Hubbard interaction governs the primary Mott physics, it is insufficient to fully explain the distinct correlation features observed between different layers and the emergence of fractional filling states. To account for these phenomena, we consider the extended moiré Hubbard model[16,17,21,33,34], where the correlation physics is governed by the competition between the on-site Hubbard repulsion, $U$, and the nearest-neighbor inter-site Coulomb interaction, $V_1$. As established by our measurements, the effective Hubbard interaction is substantially smaller in $WS_2$ than in $MoSe_2$ ($U_W < U_M$). Consequently, the effective reduction in $U$ in the $WS_2$ layer leads to an enhanced ratio of $V_1/U$, which intrinsically favors the stabilization of fractional filling states[16,35].

To further investigate the nature of these generalized Wigner crystal and stripe-like phases, we performed temperature-dependent differential reflectance measurements at $E = 0.076\,V/nm$.

Figure 4f shows the evolution of the normalized differential reflectance minimal of $X_{M1}$ signal as a function of temperature and carrier density, which highlights the spectral features associated with the fractional filling states. As the temperature increases, these features gradually weaken and eventually disappear, indicating the thermal melting of the correlated charge-ordered phases. In contrast, the integer filling states remain robust at significantly higher temperatures, reflecting the much larger energy scale of the on-site Hubbard $U$.

In Figures 4g-h, we extract the temperature dependence of the normalized differential reflectance minimal of $X_{M1}$ signal at fillings of 1/3, 1/2, and 2/3, and propose their real-space charge configurations. We find that the 1/3 and 2/3 states remain robust up to higher temperatures than the 1/2 state, with approximate melting temperatures of $T_{1/3} \sim T_{2/3} \approx 35\ K$ and $T_{1/2} \approx 25\ K$. The comparatively reduced thermal stability of the 1/2 state is consistent with the geometric frustration associated with half filling on a triangular moiré lattice. Our results, including the relative thermal stabilities and the proposed real-space configurations, are broadly consistent with previous theoretical and experimental studies of generalized Wigner crystal and stripe phases in triangular TMD moiré systems[17,19,21,23,32]. Moving forward, developing a full theoretical understanding of $V_1$ across the layer-specific fractional states, as well as their interplay with magnetic order, remains an exciting open question for future investigations.

# Methods

## Device fabrication

$MoSe_2/WS_2$ heterobilayer devices were fabricated using a dry-transfer method based on a polycarbonate (PC) film-covered polydimethylsiloxane (PDMS) based stamp. Monolayer $MoSe_2$, monolayer $WS_2$, few-layer graphite and hexagonal boron nitride (hBN) were first mechanically exfoliated from bulk crystals onto $Si/SiO_2$ substrates with a 90 nm oxide layer. For Device A, the hBN thicknesses are 19.1 nm and 20.7 nm for the top and bottom hBN, respectively. We first picked up the bottom hBN dielectric and graphite bottom gate and transferred them onto $Si/SiO_2$ substrates with pre-patterned Ti/Au (5 nm/50 nm) electrodes deposited by electron-beam evaporation through a stainless-steel shadow mask. The bottom two layers were then annealed in a tube furnace with high-purity argon at 390 °C for 6 hours. Subsequently, the graphite top gate, top hBN dielectric, graphite contact, monolayer $MoSe_2$ and monolayer $WS_2$ were picked up and stacked onto the prepared bottom two layers. Monolayer $MoSe_2$ and monolayer $WS_2$ were aligned along a clean edge with either 0° or 60° twist angles during the stacking process.

## Optical measurements

All optical measurements were performed using a 4-f confocal microscope system in a closed-loop magneto-optical cryostat system (attoDRY 2100) with a base temperature of 1.65 K. Differential reflectance (DR) measurements were performed using a supercontinuum laser (YSL SC-OEM) with 5 MHz repetition rate. The laser is focused on the sample using an apochromatic cryogenic objective (attocube LT-APO/LWD/NIR, NA 0.63). All optical signals were collected by a spectrometer (HRS-500-SS) coupled to a charge-coupled device (CCD) camera (PIXIS: 400BR eXcelon). Each DR spectrum was integrated for 1 s with an incident power density of 60 nW/μm$^2$ and normalized using a background spectrum that included all layers except the TMDs.

## Determination of electric field and doping density

The electric field and doping density of the $MoSe_2/WS_2$ heterobilayer were determined by applying symmetric or antisymmetric gate voltages to the top and bottom graphite gate ($V_{tg}$ and $V_{bg}$) based on the parallel-plate capacitor model. The electric field was calibrated as $E_{hs} = \frac{V_{tg} - V_{bg}}{\frac{\varepsilon_{TMD}}{\varepsilon_{hBN}}(t_{top}+t_{bottom})+d}$ and the doping density was calibrated as $n = \frac{\varepsilon_{hBN}V_{tg}}{et_{top}} + \frac{\varepsilon_{hBN}V_{bg}}{et_{bottom}}$. Here, $t_{top}$ = 19.1 nm and $t_{bottom}$ = 20.7 nm are the thickness of top and bottom hBN dielectrics, calibrated by optical contrast, $\varepsilon_{hBN}$ = 3.9 and $\varepsilon_{TMD}$ = 7.4 are the dielectric constants for hBN and TMDs, and $d$ = 1.2 nm is the thickness of $MoSe_2/WS_2$ heterobilayer.

## Determination of moiré density and filling factor

The moiré density $n_0$ (corresponding to one electron or hole per moiré unit cell) is defined as $n_0 = \frac{1}{L_{moiré}{}^2 sin\,(\pi/3)}$ with the moiré superlattice constant $L_{moiré} = \frac{a_{MoSe2}}{\sqrt{\delta^2+\theta^2}}$. Here, $\delta = \frac{a_{MoSe2}-a_{WS2}}{a_{MoSe2}}$ is the lattice mismatch between monolayer $MoSe_2$ ($a_{MoSe2}$= 0.332 nm) and monolayer $WS_2$ ($a_{WS2}$ = 0.319 nm), and θ is the twist angle (in radians) between the two TMD layers. The filling factor $\nu = \frac{n}{n_0}$ represents the number of carriers per moiré unit cell. We first assigned filling factors $\nu$ = 1, $\nu$ = 2 and $\nu$ = 3 insulating states from the doping-dependent DR measurement, which served as a calibration scale between gate voltages and filling factor $\nu$. Our calibration is used to calculate a relative twist angle $\theta$ = 58.4°.

**First-principles calculations**

First-principles calculations were performed for a 58.4° twisted $MoSe_2/WS_2$ heterobilayer. The relaxed atomic structure was obtained using classical force field method implemented in LAMMPS[36]. Intralayer interactions were described using the Stillinger–Weber potential[37], while the interlayer interaction between the two transition-metal dichalcogenide (TMD) layers was modeled using the Kolmogorov–Crespi (KC) interlayer potential[38]. Structural relaxation was performed until the residual atomic forces smaller than $10^{-5}$ eV/Å. Electronic-structure calculations were carried out using the SIESTA package. The Perdew–Burke–Ernzerhof (PBE) exchange-correlation functional[39] was employed. A vacuum layer of 15 Å was introduced along the out-of-plane direction to eliminate spurious interactions between periodic images. Self-consistent DFT calculations were first performed at the Γ point to get the converged charge density and the ground-state electronic structure was then obtained on a 3 × 3 × 1 Monkhorst–Pack k-point grid for the further excited states calculation. A double-ζ polarized (DZP) numerical atomic orbital basis set was employed together with a real-space grid corresponding to a mesh cutoff of 300 Ry. Excited-state properties were calculated within the GW-BSE framework using BerkeleyGW[40]. To account for quasiparticle corrections, a rigid scissors correction of 0.6 eV was applied to the conduction bands. The static dielectric matrix was obtained by folding the primitive-cell polarizability into the moiré cell representation[41], while all other quantities entering the BSE Hamiltonian were calculated explicitly in the moiré cell. A dielectric-matrix cutoff of 3 Ry and a bare Coulomb cutoff of 20 Ry were employed in constructing the electron–hole interaction kernel. The excitonic Hamiltonian was constructed using 62 valence bands and 36 conduction bands, which was sufficient to ensure convergence of the low-energy excitonic spectrum and to explicitly include the folded $MoSe_2$-derived states near the K valley arising from the moiré reconstruction. Spin-orbit coupling was included in all electronic-structure calculations but was neglected in the BSE calculations. Since this work focuses on intralayer excitons, the absence of SOC is not expected to affect the qualitative conclusions.

**Critical Field Extraction**

To determine the critical electric field, we tracked the minimum reflection signal within a narrow energy window for each given carrier filling. The magnitude of this minimum reflects the relative oscillator strength of the corresponding exciton resonance. We defined the threshold for the critical electric field to be where the relative oscillator strengths of two distinct excitonic resonances concurrently either increase or decay to 50% of their respective maximum intensity values.

**Temperature-Dependent Exciton Signal Normalization**

To quantitatively analyze the temperature-dependent optical response, we normalized the excitonic signal by fitting a polynomial baseline to the minimum optical response across the gate voltage sweep within a selected energy window. Subtracting this baseline removes the underlying background changes of the exciton resonance caused by the carrier doping density. This background removal allows us to track the anomalous intensity modulation of the excitonic features at specific filling factors as a function of temperature.

## Data availability

The main data that support the findings of this study are available within the article. More supporting data are available from the corresponding authors upon request.

## References

1. Jin, C. *et al.* Observation of moiré excitons in WSe2/WS2 heterostructure superlattices. *Nature* **567**, 76–80 (2019).
2. Seyler, K. L. *et al.* Signatures of moiré-trapped valley excitons in MoSe2/WSe2 heterobilayers. *Nature* **567**, 66–70 (2019).
3. Park, H. *et al.* Dipole ladders with large Hubbard interaction in a moiré exciton lattice. *Nat. Phys.* **19**, 1286–1292 (2023).
4. Regan, E. C. *et al.* Emerging exciton physics in transition metal dichalcogenide heterobilayers. *Nat Rev Mater* **7**, 778–795 (2022).
5. Guo, J. *et al.* Moiré-controllable exciton localization and dynamics through spatially-modulated inter- and intralayer excitons in a MoSe2/WS2 heterobilayer. *Nat Commun* **16**, 11257 (2025).
6. Naik, M. H. *et al.* Intralayer charge-transfer moiré excitons in van der Waals superlattices. *Nature* **609**, 52–57 (2022).
7. Wang, X. *et al.* Intercell moiré exciton complexes in electron lattices. *Nat. Mater.* **22**, 599–604 (2023).
8. Kistner-Morris, J. *et al.* Electric-field tunable Type-I to Type-II band alignment transition in MoSe2/WS2 heterobilayers. *Nat Commun* **15**, 4075 (2024).
9. Jauregui, L. A. *et al.* Electrical control of interlayer exciton dynamics in atomically thin heterostructures. *Science* **366**, 870–875 (2019).
10. Tang, Y. *et al.* Tuning layer-hybridized moiré excitons by the quantum-confined Stark effect. *Nat. Nanotechnol.* **16**, 52–57 (2021).
11. Ruiz-Tijerina, D. A. & Fal'ko, V. I. Interlayer hybridization and moiré superlattice minibands for electrons and excitons in heterobilayers of transition-metal dichalcogenides. *Phys. Rev. B* **99**, 125424 (2019).
12. Polovnikov, B. *et al.* Field-induced hybridization of moiré excitons in MoSe$_2$/WS$_2$ heterobilayers. *Phys. Rev. Lett.* **132**, 076902 (2024).
13. Barré, E. *et al.* Optical absorption of interlayer excitons in transition-metal dichalcogenide heterostructures. *Science* **376**, 406–410 (2022).
14. Evrard, B. *et al.* ac Stark Spectroscopy of Interactions between Moiré Excitons and Polarons. *Phys. Rev. X* **15**, 021002 (2025).
15. Qi, R. *et al.* Thermodynamic behavior of correlated electron-hole fluids in van der Waals heterostructures. *Nat Commun* **14**, 8264 (2023).
16. Morales-Durán, N., Hu, N. C., Potasz, P. & MacDonald, A. H. Nonlocal Interactions in Moiré Hubbard Systems. *Phys. Rev. Lett.* **128**, 217202 (2022).
17. Ciorciaro, L. *et al.* Kinetic magnetism in triangular moiré materials. *Nature* **623**, 509–513 (2023).
18. Qi, R. *et al.* Perfect Coulomb drag and exciton transport in an excitonic insulator. *Science* **388**, 278–283 (2025).
19. Regan, E. C. *et al.* Mott and generalized Wigner crystal states in WSe2/WS2 moiré superlattices. *Nature* **579**, 359–363 (2020).
20. Tang, Y. *et al.* Simulation of Hubbard model physics in WSe2/WS2 moiré superlattices. *Nature* **579**, 353–358 (2020).
21. Xu, Y. *et al.* Correlated insulating states at fractional fillings of moiré superlattices. *Nature* **587**, 214–218 (2020).

22. Huang, X. *et al.* Correlated insulating states at fractional fillings of the WS2/WSe2 moiré lattice. *Nat. Phys.* **17**, 715–719 (2021).
23. Tang, Y. *et al.* Dielectric catastrophe at the Wigner-Mott transition in a moiré superlattice. *Nat Commun* **13**, 4271 (2022).
24. Polovnikov, B. *et al.* Implementation of the bilayer Hubbard model in a moiré heterostructure. Preprint at https://doi.org/10.48550/arXiv.2404.05494 (2024).
25. Han, Z. *et al.* Topological Kondo insulator in MoTe2/WSe2 moiré bilayers. *Nat. Phys.* https://doi.org/10.1038/s41567-026-03170-1 (2026) doi:10.1038/s41567-026-03170-1.
26. Scherzer, J. *et al.* Correlated magnetism of moiré exciton-polaritons on a triangular electron-spin lattice. Preprint at https://doi.org/10.48550/arXiv.2405.12698 (2024).
27. Lu, Z. *et al.* Nature of Emergent Moiré Excitations in $MoSe_2$ /$WS_2$ Moiré Superlattices. *Nano Lett.* **26**, 4096–4102 (2026).
28. Gu, J. *et al.* Remote imprinting of moiré lattices. *Nat. Mater.* **23**, 219–223 (2024).
29. Guo, X. *et al.* Mott Insulator Polariton in a MoSe2/WS2 Moiré Lattice. *Nano Lett.* **25**, 15680–15685 (2025).
30. Cai, J. *et al.* Signatures of fractional quantum anomalous Hall states in twisted MoTe2. *Nature* **622**, 63–68 (2023).
31. Liu, E. *et al.* Exciton-polaron Rydberg states in monolayer MoSe2 and WSe2. *Nat Commun* **12**, 6131 (2021).
32. Jin, C. *et al.* Stripe phases in WSe2/WS2 moiré superlattices. *Nat. Mater.* **20**, 940–944 (2021).
33. Pan, H., Wu, F. & Das Sarma, S. Band topology, Hubbard model, Heisenberg model, and Dzyaloshinskii-Moriya interaction in twisted bilayer WSe 2. *Phys. Rev. Research* **2**, 033087 (2020).
34. Li, Q. *et al.* Tunable quantum criticalities in an isospin extended Hubbard model simulator. *Nature* **609**, 479–484 (2022).
35. Zhou, Y., Sheng, D. N. & Kim, E.-A. Quantum Melting of Generalized Wigner Crystals in Transition Metal Dichalcogenide Moiré Systems. *Phys. Rev. Lett.* **133**, 156501 (2024).
36. Thompson, A. P. *et al.* LAMMPS - a flexible simulation tool for particle-based materials modeling at the atomic, meso, and continuum scales. *Computer Physics Communications* **271**, 108171 (2022).
37. Jiang, J.-W. Parametrization of Stillinger–Weber potential based on valence force field model: application to single-layer $MoS_2$ and black phosphorus. *Nanotechnology* **26**, 315706 (2015).
38. Naik, M. H., Maity, I., Maiti, P. K. & Jain, M. Kolmogorov–Crespi Potential For Multilayer Transition-Metal Dichalcogenides: Capturing Structural Transformations in Moiré Superlattices. *J. Phys. Chem. C* **123**, 9770–9778 (2019).
39. Perdew, J. P., Burke, K. & Ernzerhof, M. Generalized Gradient Approximation Made Simple. *Phys. Rev. Lett.* **77**, 3865–3868 (1996).
40. Deslippe, J. *et al.* BerkeleyGW: A massively parallel computer package for the calculation of the quasiparticle and optical properties of materials and nanostructures. *Computer Physics Communications* **183**, 1269–1289 (2012).
41. You, J.-Y. *et al.* Moiré excitons in generalized Wigner crystals. *Proc. Natl. Acad. Sci. U.S.A.* **123**, e2531259123 (2026).

## Acknowledgements

A.Y.J., H.Y., and Q.L. acknowledge support from the donors of the ACS Petroleum Research Fund under Doctoral New Investigator (DNI) Grant #68101-DNI5. A.Y.J. also acknowledges support from the University of California, Riverside (UCR) Academic Senate Regents Faculty Fellowship and UCR start-up funds. C.-E. H. acknowledges the funding support of the National Science and Technology Council, Taiwan, under the Grant No. 113-2917-I-032001. Z.L. acknowledges the support from United States Department of Energy (DOE), Office of Basic Energy Sciences, under the Contract No. DE-SC0026336 during the revision stage of this work. C.-E.H. and H.-C.H. acknowledge the support by National Science and Technology Council under Grant Nos. 113-2112-M-032-013, 114-2112-M-032-009-MY3 and 114-281-M-032-012-MY2. K.W. and T.T. acknowledge support from the CREST (JPMJCR24A5), JST and World Premier International Research Center Initiative (WPI), MEXT, Japan.

## Author contributions

A.Y.J conceived the research. H.Y. fabricated the device with the help of Q.L. H.Y. performed the optical measurements. H.Y., Q.L. and A.Y.J analyzed the data. C.-E.H., H.-C.H., and Z.L. performed and analyzed first-principles DFT and GW-BSE calculations. K.W. and T.T. grew the hBN crystals. All authors discussed the results and wrote the manuscript.

## Competing interests

The authors declare no competing interests.

## Figures

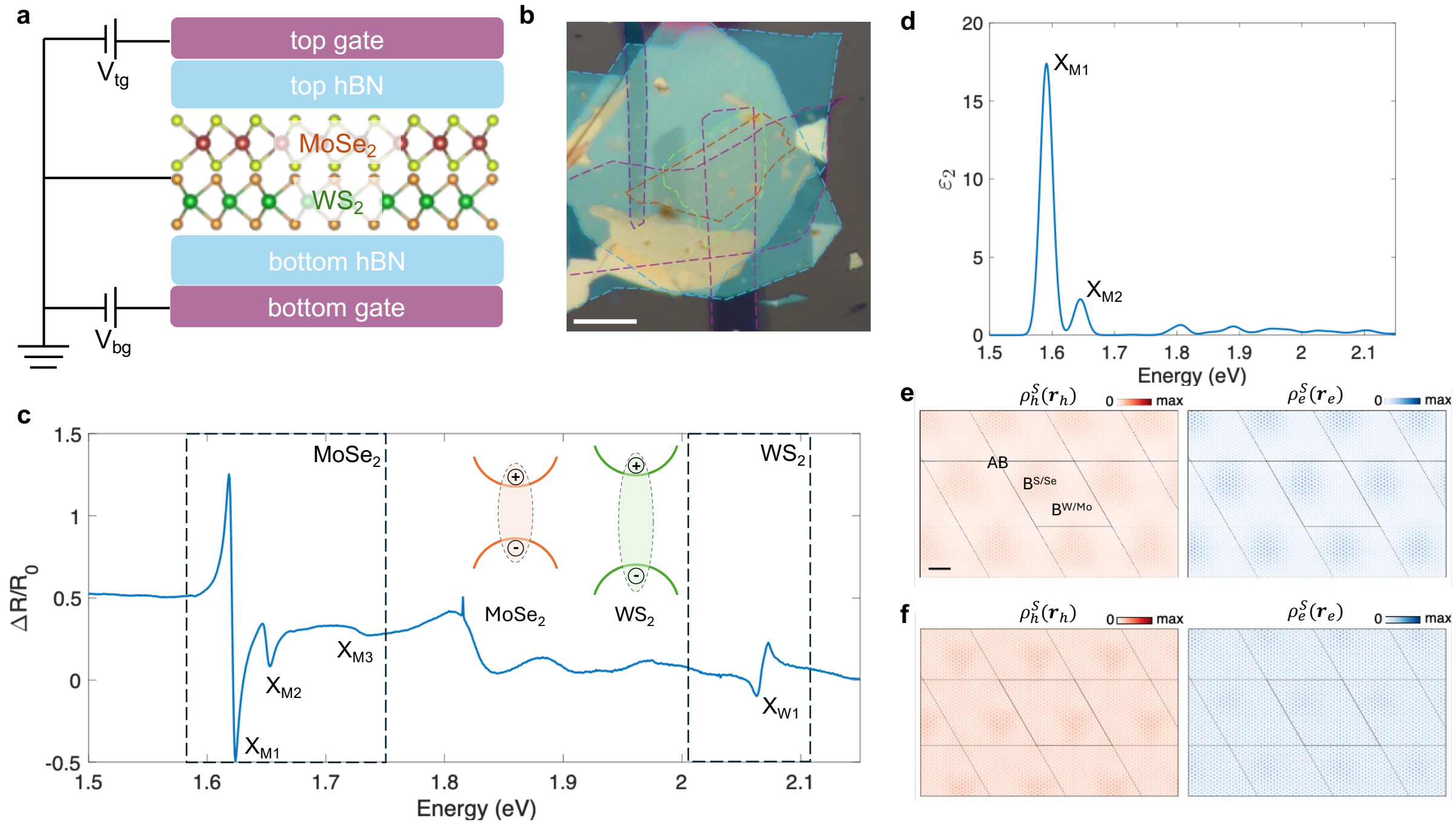

**Fig. 1 Moiré exciton species in H-stacking $MoSe_2/WS_2$ moiré superlattice. a.** Schematic of the dual-gated $MoSe_2/WS_2$ moiré superlattice device. **b.** Optical image of the device. Scale bar: 10 µm. The $MoSe_2$, $WS_2$, graphite gates and contact graphite region are indicated by orange, green, light purple and deep purple dashed outlines. **c.** Differential reflectance spectrum measured at zero carrier doping and electric field. The energy window of $MoSe_2$ and $WS_2$ are identified with dashed rectangular respectively. The inset shows the initial band alignment of $MoSe_2$ and $WS_2$. **d.** Theoretically calculated absorption spectrum features two resonance peaks of $MoSe_2$ using the GW-BSE method. **e-f**. Corresponding spatial localization of exciton $X_{M1}$ (**e**) and $X_{M2}$ (**f**) in the moiré sites obtained from the theoretical calculations. The excited hole distribution is shown on the left and the excited electron distribution on the right. Scale bar: 2 µm.

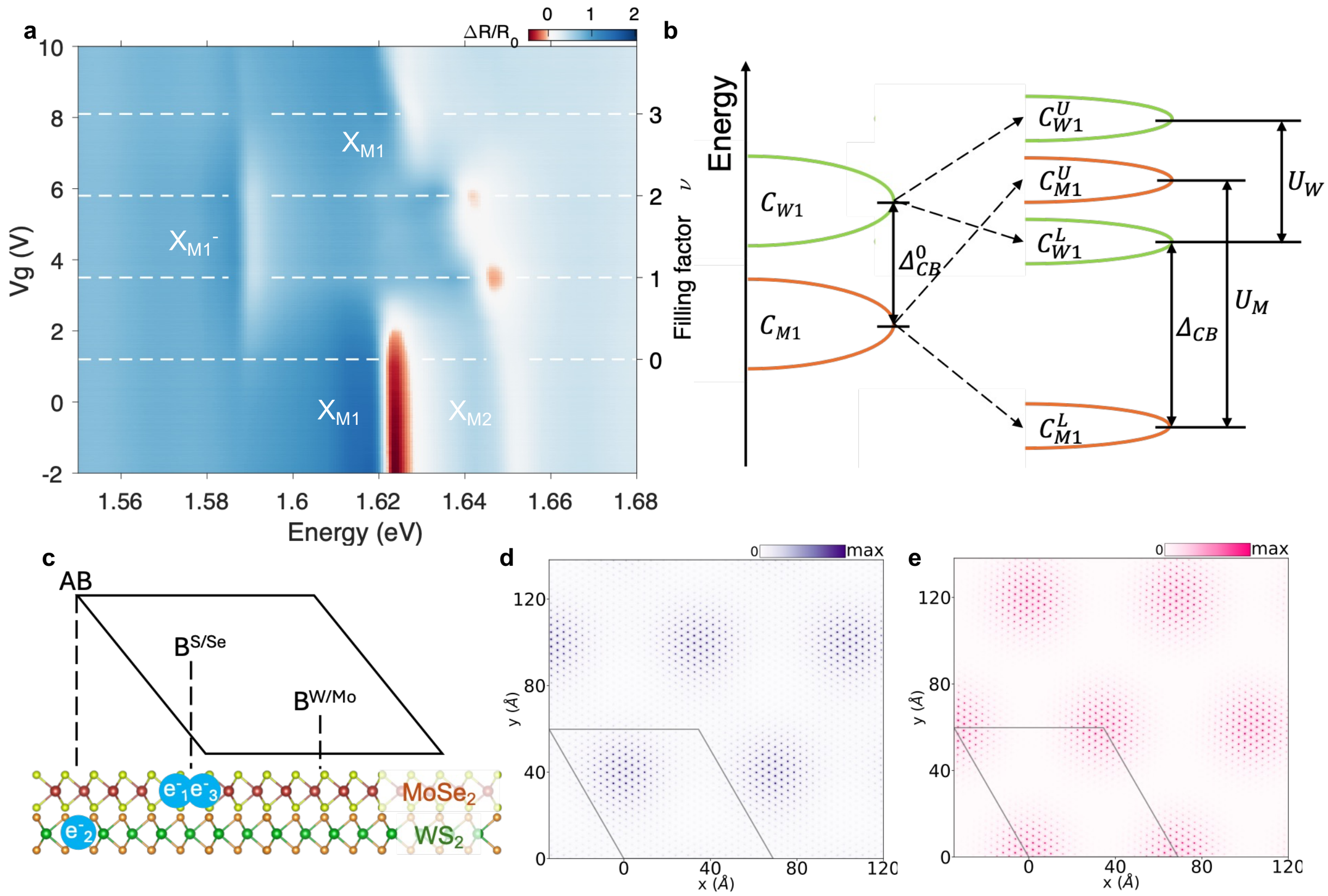

**Fig. 2 Moiré exciton behavior as the change of filling factor and relative band configuration.** **a.** Doping-dependent differential reflectance in the $MoSe_2$ A-exciton energy window at zero electric field. $X_{M1}$ and $X_{M2}$ are identified as moiré excitons. The feature near 1.59 eV is assigned to be the trion of $X_{M1}$. **b.** Band alignment of $MoSe_2$ and $WS_2$ without (left) and with (right) the inclusion of the Hubbard $U$. The original conduction band of $MoSe_2$ ($WS_2$) splits to an upper Hubbard band $C_{M1}^{U}$ ($C_{W1}^{U}$) and lower Hubbard band $C_{M1}^{L}$ ($C_{W1}^{L}$) due to the on-site Coulomb interaction $U_M$ ($U_W$). **c.** The spatial distribution of electrons upon sequential filling of the first, second, and third electrons in the moiré superlattice. **d-e.** Real-space charge density distributions of the lowest-energy conduction-band states from the K valley of $MoSe_2$ (**d**) and $WS_2$ (**e**) obtained from first-principles DFT calculations.

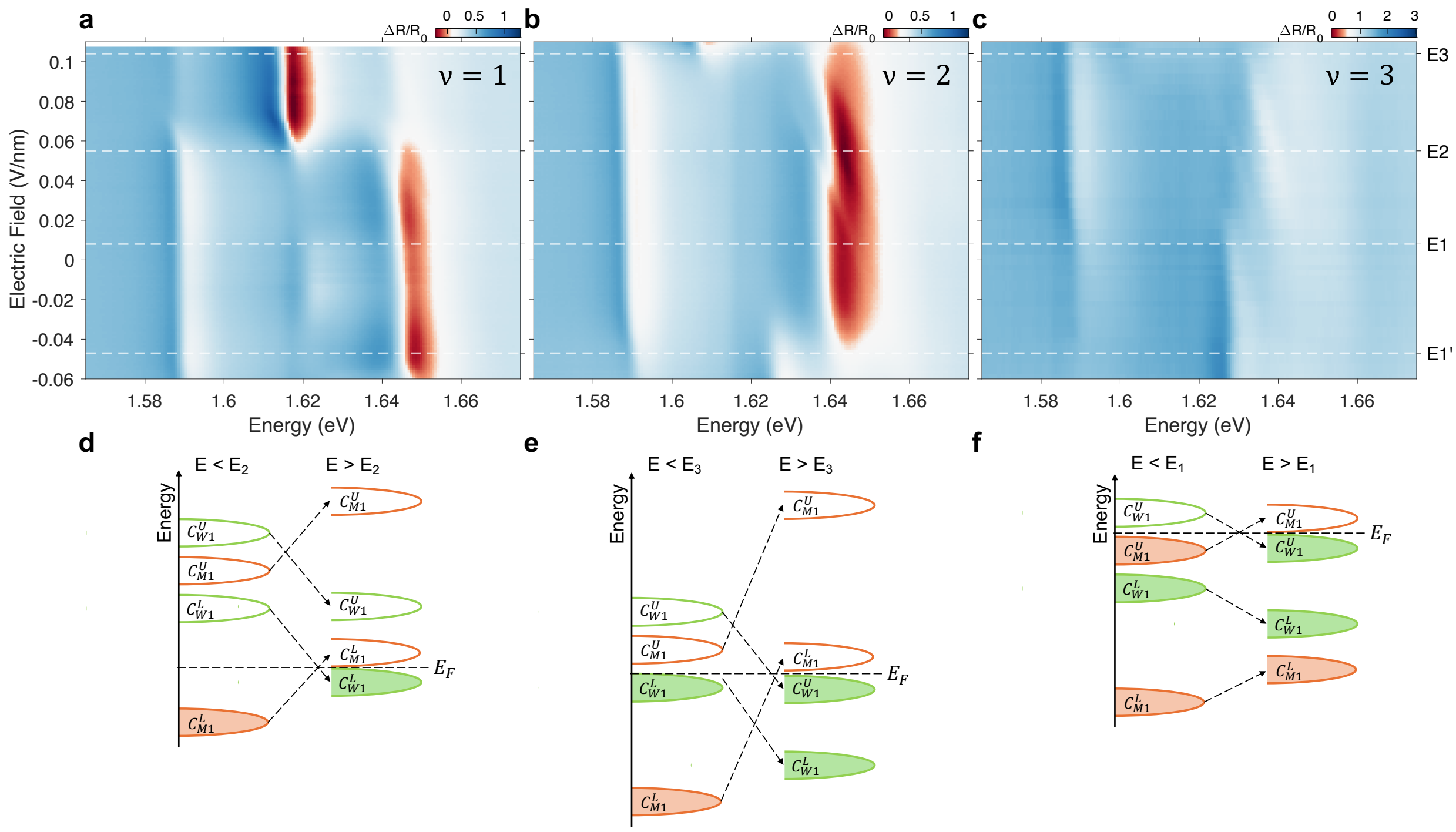

**Fig. 3 Electric-field tunable moiré excitons species and Hubbard band between two layers. a-c.** Electric-field dependent differential reflectance in the energy window of $X_{M1}^{-}$, $X_{M1}$ and $X_{M2}$, at fixed filling factor ν =1 (**a**), ν = 2 (**b**) and ν = 3 (**c**). Critical electric fields marking the switch between Hubbard band ordering are labeled as $E_{1'}$ = -0.047V/nm, $E_1$ = 0.008V/nm, $E_2$ = 0.055V/nm and $E_3$ = 0.104V/nm. **d-f.** Evolution of the Hubbard band alignment at filling factors ν = 1 (**d**), ν = 2 (**e**) and ν = 3 (**f**) in the vicinity of critical electric field $E_2$, $E_3$ and $E_1$.

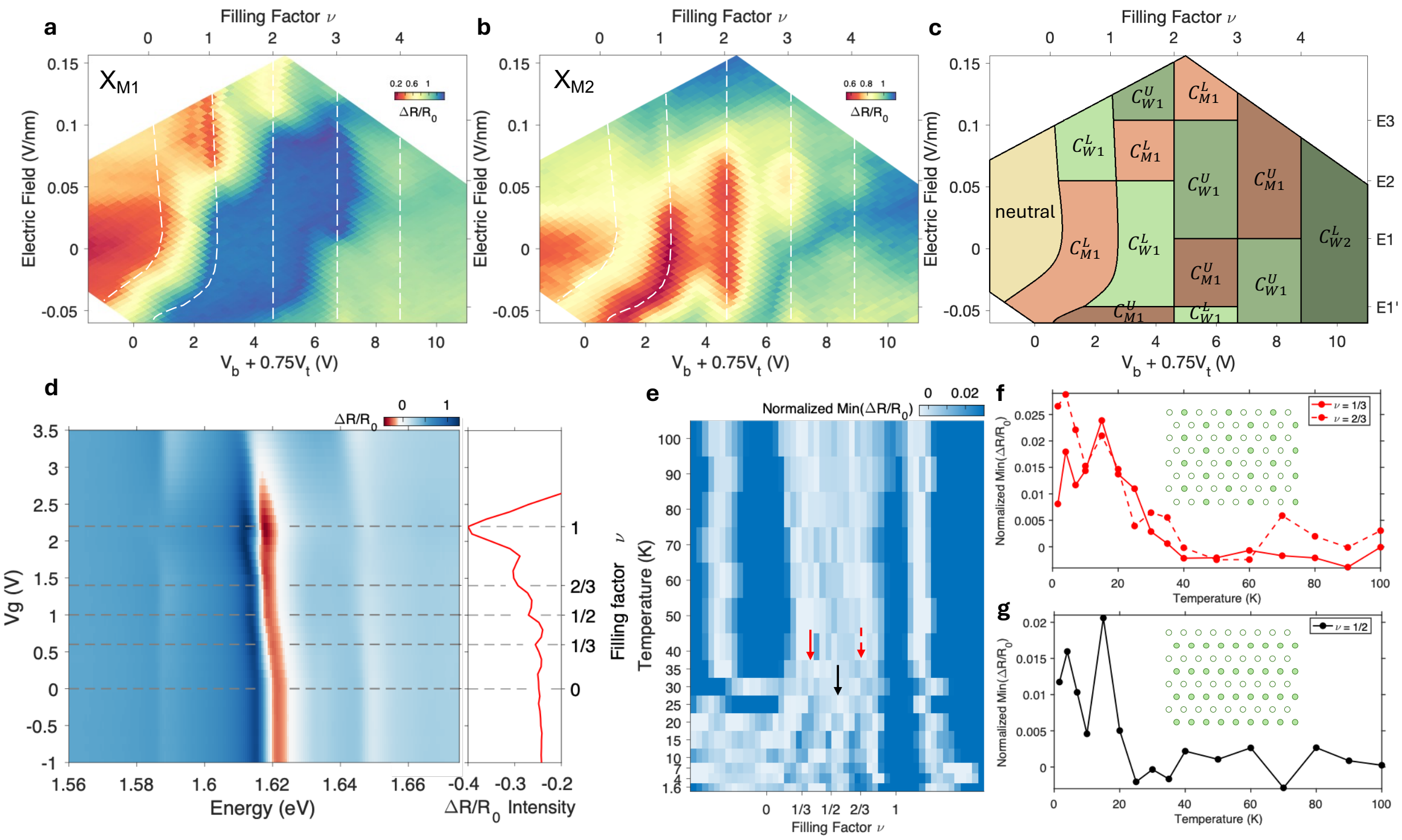

**Fig. 4 Phase diagram and electric-field induced correlated states. a-b.** The differential reflectance mapping of moiré exciton $X_{M1}$ (**a**) and $X_{M2}$ (**b**) with the change of electric field and filling factors. **c.** Electron-filled Hubbard band evolution phase diagram with the change of electric field and filling factors. **d.** Doping dependent differential reflectance of $X_{M1}$ under electric field E = 0.076V/nm, with the intensity minimum emerging at fractional filling factors. **e**. Temperature dependence of the normalized differential reflectance minimal of the $X_{M1}$ signal as a function of the filling factors. **f-g**. Temperature dependence of the normalized differential reflectance minimal and proposed real-space charge configurations for filling factors 1/3 and 2/3 (**f**) and 1/2 (**g**).

## Extended Data

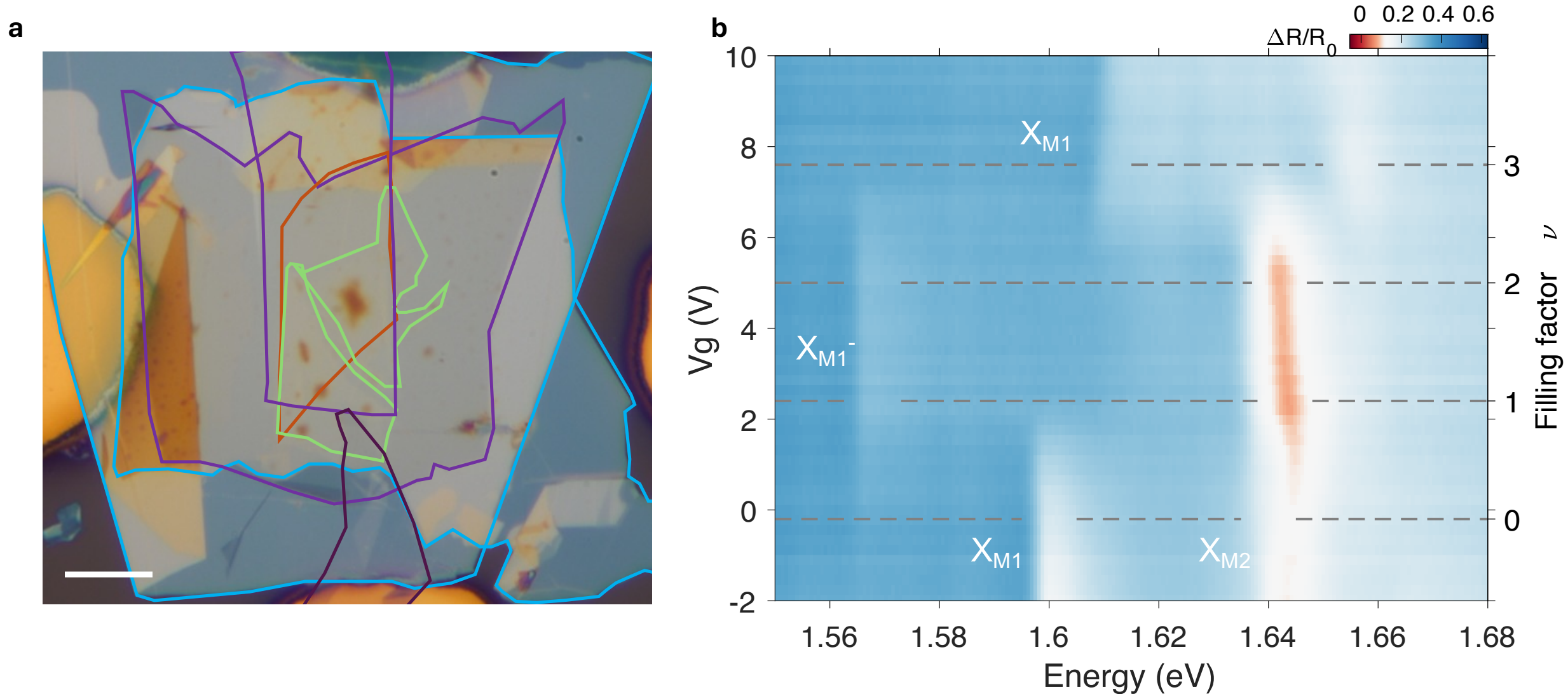

**Extended Data Fig. 1 Additional results from Device B**. **a.** Optical image of Device B with a calibrated twist angle $\theta \sim 58.9$ degrees. The $MoSe_2$, $WS_2$, graphite gates and contact graphite region are indicated by orange, green, light purple and deep purple dashed outlines. **b.** Doping-dependent differential reflectance of Device B in the $MoSe_2$ A-exciton energy window at zero electric field. $X_{M1}$ and $X_{M2}$ are identified as moiré excitons.

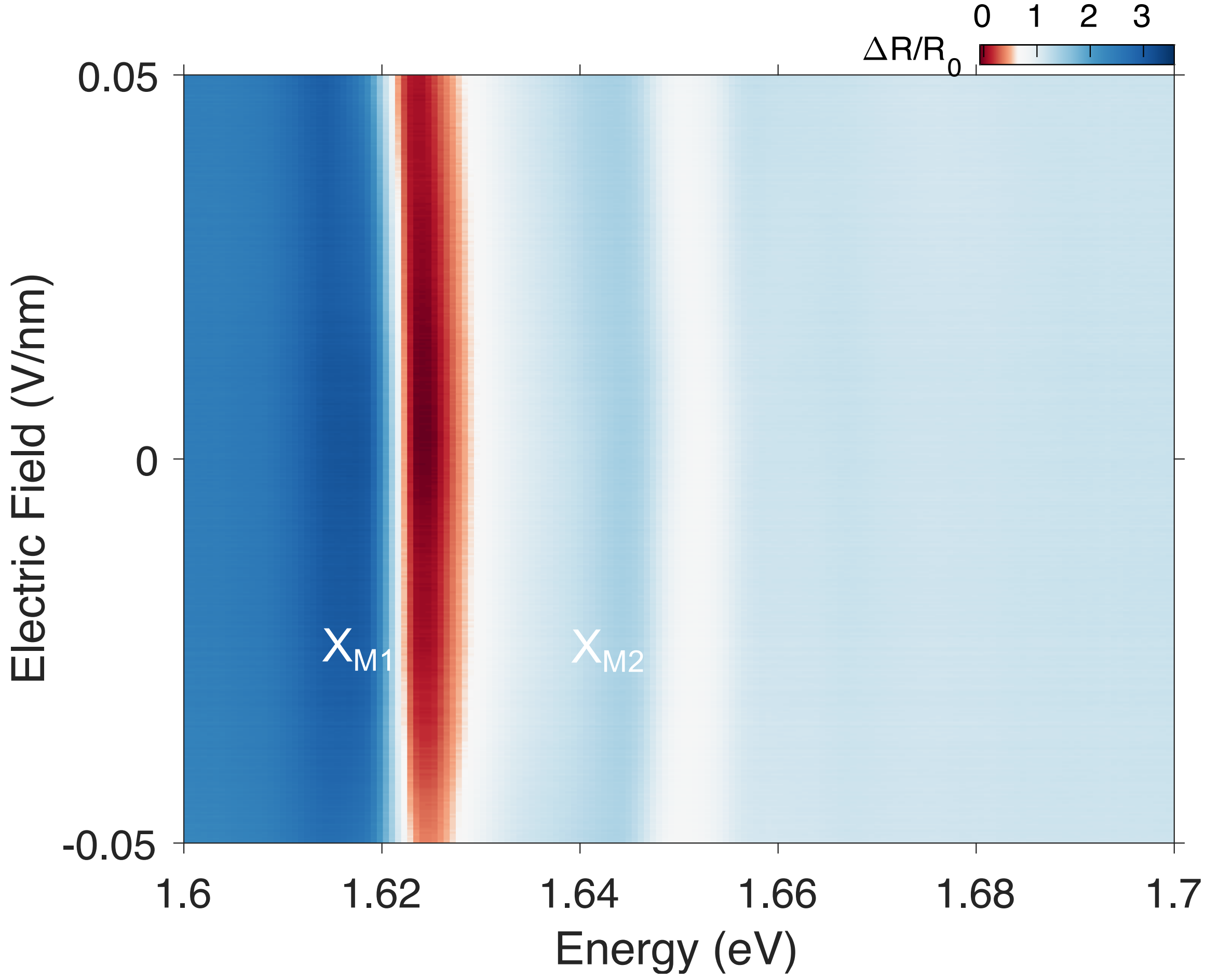

**Extended Data Fig. 2 Intralayer character of moiré excitons**. Electric field-dependent differential reflectance of Device A in the $MoSe_2$ A-exciton energy window at zero carrier doping. $X_{M1}$ and $X_{M2}$ do not shift with an electric field, confirming their intralayer character.

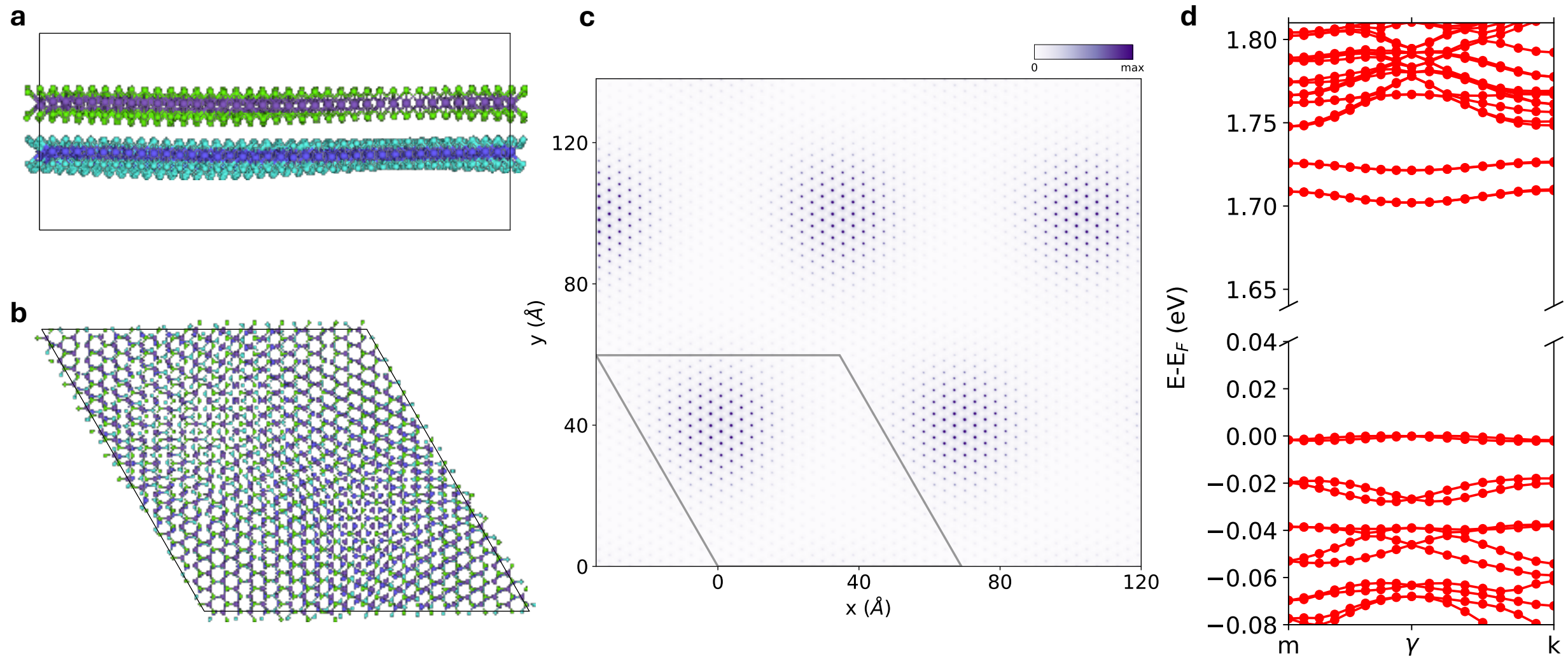

**Extended Data Fig. 3 Detailed First-Principles calculations**. **a.** Relaxed atomic structure of a 58.4° twisted $MoSe_2/WS_2$ heterobilayer using classical force fields. Side view. **b.** Top view of the same structure. **c.** Spatial distribution of the electronic charge density associated with the second conduction band of $MoSe_2$, originating from the K valley, which is also predominantly localized at the $B^{S/Se}$ site. **d.** Calculated electronic band structure of the reconstructed H-type $MoSe_2/WS_2$ moiré superlattice.

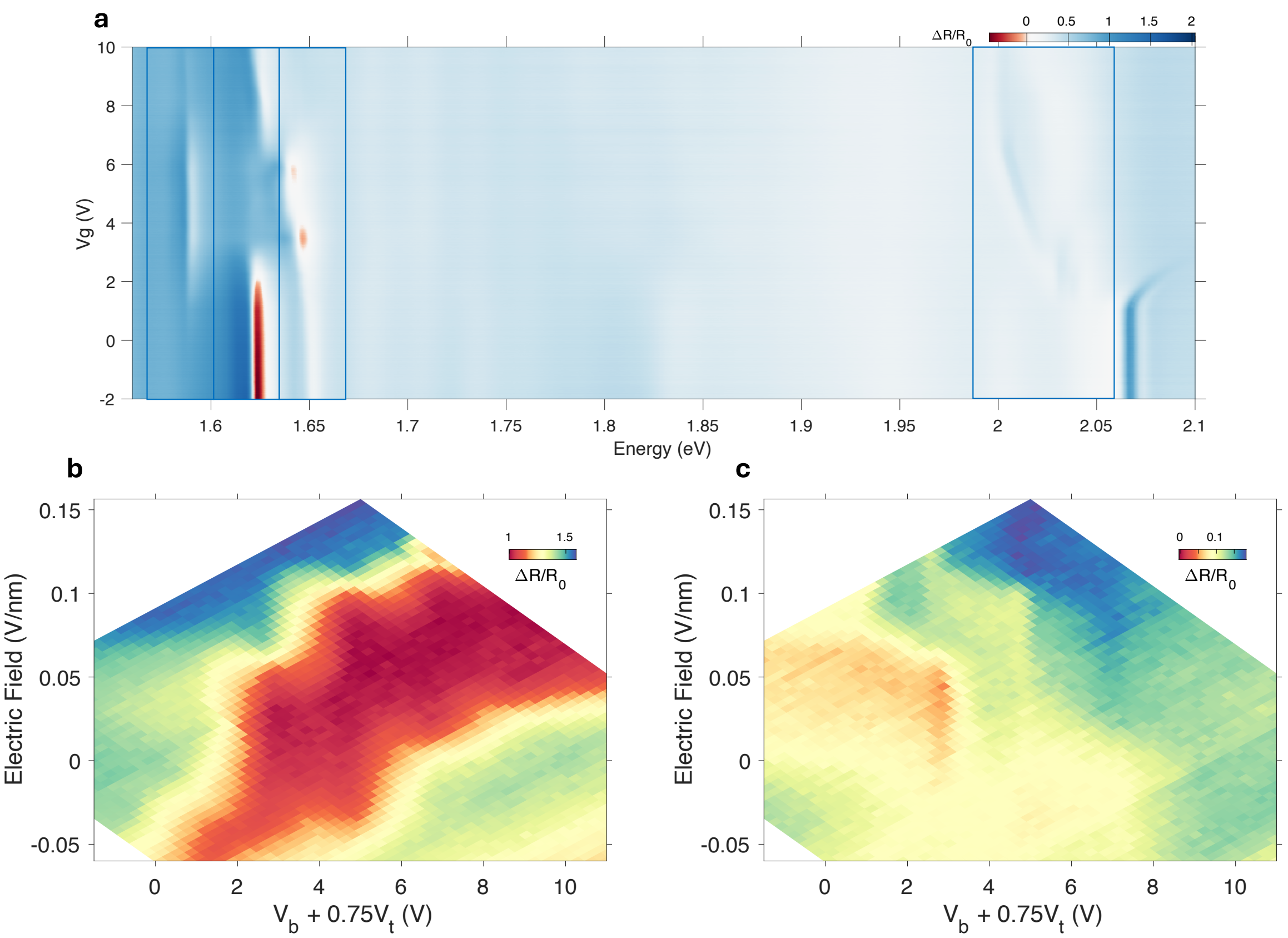

**Extended Data Fig. 4 Additional results on phase diagram**. **a.** Doping-dependent differential reflectance in both $MoSe_2$ and $WS_2$ energy windows. The tracked energy windows of $X_{M1}^-$, $X_{M1}$, $X_{M2}$ and $X_{W1}$ are identified with blue rectangles. **b-c.** The differential reflectance mapping of moiré exciton $X_{M1}^-$ (**b**) and $X_{W1}$ (**c**) with the change of electric field and filling factors.